\documentclass[useAMS,usenatbib]{mn2e}

\usepackage{amsmath,texlogos}
\usepackage{natbib}
\usepackage{subfigure}
\usepackage{graphicx}
\usepackage{txfonts}
\usepackage[T1]{fontenc}
\usepackage{aecompl}
\usepackage{times}

\title[TSARDI algorithm]{TSARDI: a Machine Learning data rejection algorithm for transiting exoplanet light curves}
\author[Mislis D.]{Mislis D.$^{1}$\thanks{E-mail:dmislis@qf.org.qa}, Pyrzas S.$^{1}$, Alsubai K.A.$^{1}$\\ \\
$^{1}$Qatar Environment and Energy Research Institute (QEERI), Hamad Bin Khalifa University (HBKU), Qatar Foundation, P.O. Box 5825, Doha, Qatar\\
}

\begin{document}

\date{Accepted ??. Received ??; in original form ??}

\pagerange{\pageref{firstpage}--\pageref{lastpage}} \pubyear{2002}

\maketitle

\label{firstpage}

\begin{abstract}
We present \texttt{TSARDI}, an efficient rejection algorithm designed to improve the transit detection efficiency in data collected by large scale surveys. \texttt{TSARDI} is based on 
the Machine Learning clustering algorithm \texttt{DBSCAN}, and its purpose is to serve as a robust and adaptable filter aiming to identify unwanted noise points left over from data 
detrending processes. \texttt{TSARDI} is an unsupervised method, which can treat each light curve individually; there is no need of previous knowledge of any other field light curves. 
We conduct a simulated transit search by injecting planets on real data obtained by the QES project and show that \texttt{TSARDI} leads to an overall transit detection efficiency 
increase of $\sim$11\%, compared to results obtained from the same sample, but using a standard sigma-clip algorithm. For the brighter end of our sample (host star magnitude < 12),
\texttt{TSARDI} achieves a detection efficiency of $\sim$80\% of injected planets. While our algorithm has been developed primarily for the field of exoplanets, it is easily adaptable 
and extendable for use in any time series.

\end{abstract}


\section{Introduction}
Large-scale, ground-based surveys for transiting extrasolar planets (e.g. HAT: \citealt{hatn}; TreS: \citealt{tres}; SuperWASP: \citealt{swasp}; KELT: \citealt{kelt}; 
QES: \citealt{alsubai13}) have been the steady work-horses of the field during the last 15 years. Almost since the beginning of these endeavours, it became readily apparent that the 
data collected were severely affected by \emph{systematics}, i.e. unwanted flux variations introduced by fixed, ordered trends, such as airmass and seeing variations, 
colour-dependent extinction, object merging etc.

The answer to the problem came with the development of \emph{detrending} algorithms, with \texttt{TFA} \citep{tfa} and \texttt{SysRem} \citep{tamuz} being among the most well-known.
While both these two, as well as other similar detrending algorithms (e.g. \citealt{mislis10,ofir,still,mislis17}), can effectively remove (or at least minimise) the effects of major
trends, they are not necessarily designed to tackle more subtle data irregularities that remain after detrending. Such irregularities can arise from infrequent and/or aperiodic
events, e.g. the presence of cirrus, variations in atmospheric transparency and the presence of dust, variations in the sky background etc. 

By design, large-scale surveys carry out long campaigns, observing their fields for a given time-period (mainly defined by the field's visibility in the sky) and returning to them 
when next visible; as such, field observations can span years, with considerable time gaps inbetween. Additionally, it is not uncommon for surveys to combine observations from 
different stations in a multi-longitude mode of observing. The longer a campaign lasts and the more data from different years and/or places are combined, the more susceptible light 
curves become to the irregular variations described above. 

The net effect of these variations is mainly two-fold: (i) randomly distributed nights with higher RMS than the majority and (ii) in the absence of global flux calibration, nights 
with a mean flux level distinctly different from the overall light curve mean. While individually (i.e. from a single night) the effect on the overall light curve is most likely 
negligible, it can quickly escalate with additional nights and conceivably reach the 1\% level of a typical transit; the end-result, when phase-folding the data to look for periodic 
transit signals, is a ``puffed up'' light curve, i.e. a light curve with an RMS higher than it \emph{should} have, which can prove detrimental in identifying transit signals. 

In recent years, Machine Learning (ML) algorithms have started becoming popular in a variety of research topics in Astrophysics, with the field of exoplanets prominent among them
(e.g. \citealt{torniainen,carrasco,masci,armstrongb,mccauliff,mislis16,armstrongc,armstrongd}). In this paper, we make use of ML, and develop a filtering algorithm, designed to tackle 
data irregularities that remain after the detrending process.

We present the \emph{\texttt{TSARDI}} (TimeSeries Analysis for Residual Data Irregularities) algorithm; our approach is generally based on the \emph{class identification} 
methodology, i.e. the goal is to group the data points of a light curve into meaningful subclasses. To achieve this, we use the \emph{clustering} algorithm \texttt{DBSCAN}
(Density-Based Spatial Clustering of Applications with Noise), originally developed by \cite{DBSCAN}, as a more efficient and more effective way of discovering clusters of arbitrary 
shape, compared to other clustering algorithms (e.g. CLARANS and \textit{k-means}/\textit{k-medoid} partitioning algorithms). \texttt{DBSCAN} has been used by the K2 mission in order to 
optimize the photometric aperture size \citep{barros} and by \textit{ASTErIsM} \citep{tramacere} for galaxy detection and shape classification, but has otherwise attracted little 
attention in astronomical applications.

In Section\,\ref{sec:algor} we describe the algorithm; Section\,\ref{sec:results} illustrates the effect of the algorithm on detrended light curves and describes the results of our 
simulated transit search; and Section\,\ref{sec:conc} contains some concluding remarks.

\section{The algorithm}
\label{sec:algor}

In what follows, we will first give a brief overview of the \texttt{DBSCAN} algorithm and summarise the necessary definitions; interested readers are referred to \cite{DBSCAN} for the complete, 
in-depth analysis. Subsequently, we will give a detailed description of \texttt{TSARDI}. We note here that \texttt{TSARDI} was built upon the \texttt{DBSCAN} routines as implemented 
in Python's \texttt{scikit-learn} package\footnote{http://scikit-learn.org} \citep{pedregosa}.

\subsection{DBSCAN}
\label{subsec:dbs}

The main function of \texttt{DBSCAN} is to take a \emph{sample} of points $S$ (in this case, a light curve) and organise all points in $S$ into \emph{clusters}, $C$. 
To achieve this, \texttt{DBSCAN} defines (i) a \emph{distance function}\footnote{This can be any appropriate function of a given problem.}, $dist(p,q)$, between points $p,q\in S$; (ii) an 
upper-limit/maximum value for the distance function, denoted as \emph{Eps}; and (iii) a minimum number of points \emph{MinPts}. We can now proceed to the following definitions: \\

\noindent
The Eps-neighbourhood of a point p, $N_{\rm{Eps}}(p)$, is given by $N_{\rm{Eps}}(p)\,=\,\{ q\in S \,|\, dist(p,q) \leq \rm{Eps}\}$ \\

\noindent
A point $p$ is directly density-reachable from another point $q$, if the following two conditions are met: (1) $p \in N_{\rm{Eps}}(q)$ and (2) 
$|N_{\rm{Eps}}(q)|\,\geqslant MinPts$. In this case, i.e. when $|N_{\rm{Eps}}(q)|\,\geqslant MinPts$, $q$ is called a \emph{core-point}. \\

\noindent
If there is a chain of points $p_{1}=q,\ldots,p_{n}=p$ such that $p_{i+1}$ is directly density-reachable from $p_{i}$, then $p$ is density-reachable from $q$.\\

\noindent
For three points $p,q,w$, if both $p$ and $q$ are density-reachable from $w$, then $p$ and $q$ are density-connected. \\

With the above definitions, we can define a \emph{cluster} $C$ as a non-empty subset of $S$, satisfying the conditions: (I) if $p\in C$ and $q$ is density-reachable from $p$, then
$q \in C$; and (II) $\forall p,q \in C$: $p$ is density-connected to $q$. Finally, we note that points which do not belong to any cluster, are considered \emph{noise points}.

\begin{figure}
 \includegraphics[width=9cm]{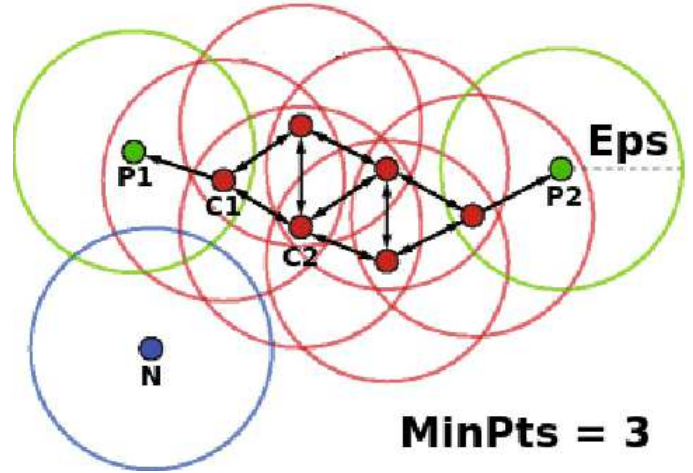}
 \caption{A schematic representation of the DBSCAN definitions. See text for details.}
 \label{fig:odbs}
\end{figure}

An overview of these definitions is visualised in Figure\,\ref{fig:odbs}. The Eps-neighbourhood of each point is shown as a circle with radius Eps (shown as a dashed line), 
while MinPts = 3. All red points are core points, because $|N_{\rm{Eps}}(q)|\,\geqslant MinPts$. Point C1 is directly density-reachable from point C2, because it belongs to the 
Eps-neighbourhood of C2 and C2 is a core point; the opposite is also true, i.e. 
point C2 is directly density-reachable from point C1. This pair-wise relation is indicated by the bidirectional arrow. Point P1 is directly density-reachable from point C1; however 
the opposite is \emph{not} true. This is indicated by the single arrow. Point P1 is density-reachable from point C2 (because C2\,$\rightarrow$\,C1\,$\rightarrow$\,P1). Points P1 and 
P2 are density-connected, as they are both density reachable from e.g. C2. \emph{All} red points together with points P1 and P2 belong to the \emph{same} cluster. Finally, point N 
does \emph{not} belong to the cluster and is considered a noise point.

It is obvious that the classification of a set into one, or more, clusters and (most importantly) the identification of those points that do not belong to \emph{any} cluster, depends 
heavily upon the choice of the values for Eps and MinPts.

\subsection{TSARDI}
\label{subsec:tsardi}
The \texttt{TSARDI} algorithm consists of two major parts: (1) the core-algorithm part and (2) the external-shell part. A detailed account of both parts follows.

\subsubsection{The core-algorithm part}
\label{subsubsec:cap}

The core-algorithm part is based on four distinct, but chain-linked steps. Each step implements \texttt{DBSCAN} with step-unique distance function and values for Eps and 
MinPts. At each step, the target is to classify the input light curve points into one or more clusters and identify the \emph{noise points}. These noise points are subsequently 
eliminated, and the resulting ``filtered'' light curve is used as input for the next step.

In what follows, we assume that our light curve consists of $N$ pairs of time-and-flux values, $P_{i}=(t_{i},f_{i})$ with $i=1,\ldots,N$ and spans a total of $K$ nights of 
observation. \\

\begin{figure}
 \includegraphics[width=9cm]{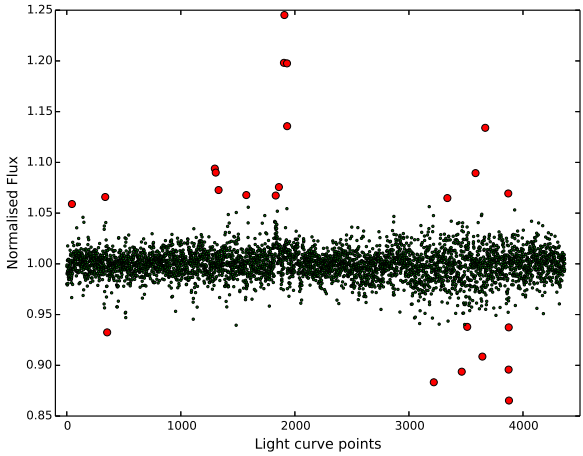}
 \caption{\textbf{STEP 1} of our algorithm: the single ``good points'' cluster (smaller, green dots) and the noise points (larger, red dots), identified while treating the light curve 
 as a whole, based on the median absolute flux difference between two consecutive points.}
 \label{fig:step_1}
\end{figure}

\noindent
\textbf{STEP 1:} The first distance function, $df_{1}$, is the absolute flux difference between two consecutive points, that is  $df_{1}(j) = abs(f_{j+1}-f_{j})$ for 
$j=1,\ldots,N-1$. Here, the light curve is treated as one ``whole'', i.e. a continuous timeseries, and ``consecutive'' is used in an ordinal sense, so that the first point of one night 
is consecutive to the last point of the previous night. As $Eps_{1}$, we set the median value of all calculated $df_{1}(j)$ times a \emph{multiplication factor} $mf_{I}$, that is
$Eps_{1}=mf_{I}\times\widetilde{df_{1}(j)}$; the function of $mf_{I}$ will become obvious in Sec.\,\ref{subsubsec:lsr} and \ref{subsubsec:esp}. $N$/100 is set as $MinPts_{1}$. 
Figure\,\ref{fig:step_1} visualises the first step. \\

\begin{figure}
 \includegraphics[width=9cm]{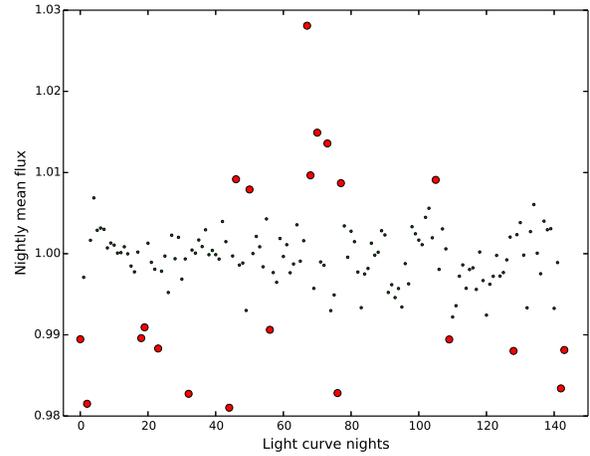}
 \caption{\textbf{STEP 2} of our algorithm: the single ``good points'' cluster (smaller, green dots) and the noise points (larger, red dots), identified after binning the light curve 
 per night, based on the mean flux value of a given night.}
 \label{fig:step_2}
\end{figure}

\noindent
\textbf{STEP 2:} We now split the light curve into its constituent nights; for each night, we calculate its mean flux value $\bar{f}_{k}$, with $k=1,2,\ldots, K$. The second 
distance function, $df_{2}$, is the absolute difference of mean fluxes between two consecutive nights\footnote{Here, again, ``consecutive'' is used in an ordinal sense.}, that is  
$df_{2}(m) = abs(\bar{f}_{m+1}-\bar{f}_{m})$ for $m=1,\ldots,K-1$. As before, $Eps_{2}$ is set as the median value of all calculated $df_{2}(m)$ values times \emph{another} 
multiplication factor $mf_N$, so $Eps_{2}=mf_N\times\widetilde{df_{2}(m)}$. As $MinPts_{2}$ we set $K$/5. Figure\,\ref{fig:step_2} visualises the second step. \\

\begin{figure}
 \includegraphics[width=9cm]{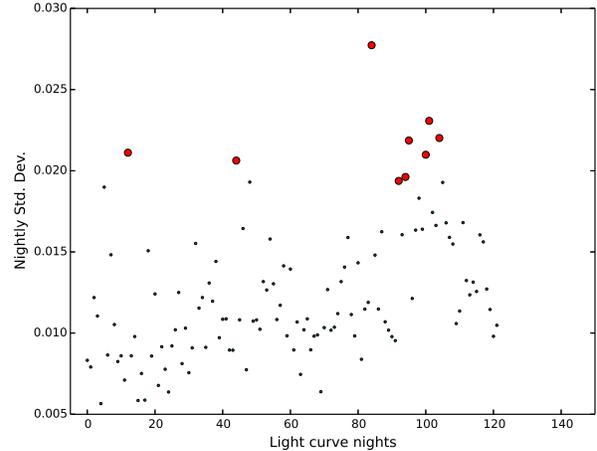}
 \caption{\textbf{STEP 3} of our algorithm: similar to Step 2, but this time based on the standard deviation of a given night. Notice how the number of nights has decreased, after
 discarding some nights in the previous step.}
 \label{fig:step_3}
\end{figure}

\noindent
\textbf{STEP 3:} This step is almost identical to the previous, only this time we calculate the \emph{standard deviation} of the flux values of a given night, $\sigma_{k}$.
The distance function, $df_{3}$, is the absolute difference of standard deviations between two consecutive nights; $Eps_{3}$ is taken to be the median value of all $df_{3}(m)$ times
$mf_N$, i.e. the \emph{same} multiplication factor as in Step 2; and $MinPts_{3}$ is again $K$/5. Figure\,\ref{fig:step_3} visualises the third step. \\

\begin{figure}
 \includegraphics[width=9cm]{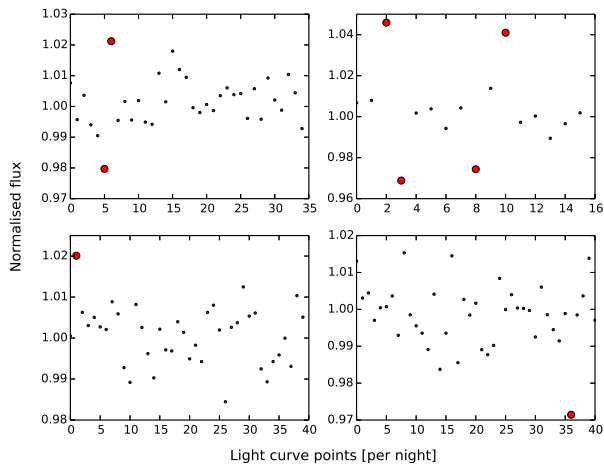}
 \caption{\textbf{STEP 4} of our algorithm: four representative examples of the output are shown. In this step, each night is treated as a ``mini-light curve'', and the identification
 of good points and noise points is based on the median absolute flux difference between two consecutive points.}
 \label{fig:step_4}
\end{figure}

\noindent
\textbf{STEP 4:} The fourth, and final, step is similar to Step 1, in that it uses the absolute flux difference between two consecutive points, but in this case, the light curve is
once more split into its constituent nights (as in Steps 2 \& 3). For a given night $k$, with $n$ points, we calculate $df^{k}_{4}(j) = abs(f_{j+1}-f_{j})$ where $j=1,\ldots,n-1$.
Both the Eps and MinPts values are set on a \emph{per night} basis, as the median of the corresponding $df^{k}_{4}(j)$ values times $mf_{I}$ (the multiplication factor of Step 1), and 
as $n$/5, respectively. Figure\,\ref{fig:step_4} visualises the fourth step.\\

\begin{figure}
 \includegraphics[width=9cm]{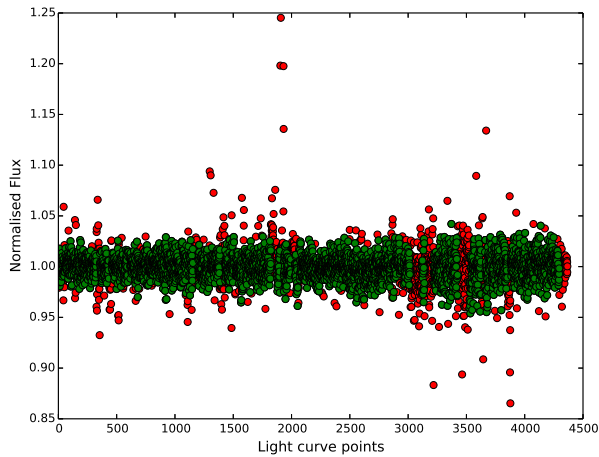}
 \caption{The original light curve (red points) and the \texttt{TSARDI}-filtered, final light curve (green points). For clarity, we plot the light curve as consecutive points, not
 according to their timestamps.}
 \label{fig:lcgen}
\end{figure}

In Figure\,\ref{fig:lcgen} we plot both the original light curve of this example, as well as the resulting \texttt{TSARDI}-filtered light curve; in terms of numbers of points, 
the original and the final light curve consist of 4,367 and 3,636 respectively, i.e. the filtered light curve retains $\sim$83\% of the original number of points.

\subsubsection{Large signals and over-rejection}
\label{subsubsec:lsr}
A common caveat of clipping/rejection algorithms is the possibility of rejecting valid points (i.e. true signal) that are found far away from the majority of points in a light 
curve, as is the case in (deeply) eclipsing binaries and even ``large planetary''-sized bodies (e.g. a brown dwarf transiting a late-K or an M-dwarf star).

\begin{figure}
 \includegraphics[width=9cm]{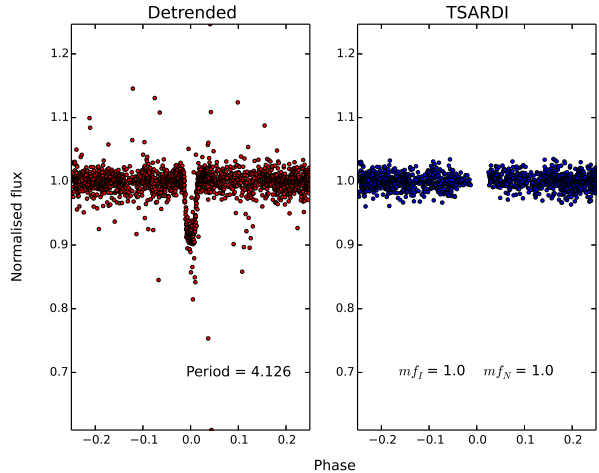}
 \caption{An example transit of a 2\,$\rm{R_{J}}$ object with an orbital period of 4.126\,d, yielding a depth of 0.132. On the left-hand side panel, the detrended light curve. Running 
 the core-algorithm part of \texttt{TSARDI} in a straightforward fashion, i.e. with $mf_{I}\,=\,mf_{N}\,=\,1$ (effectively without multiplication factors) results in the in-transit 
 points being classified as noise points and, therefore, rejected.}
 \label{fig:over_rej}
\end{figure}

For a straightforward implementation of the core-algorithm part of \texttt{TSARDI} with rigid Eps values in each Step (in other words, \emph{without} the multiplication factors 
$mf_I$ and $mf_N$) there is a high probability of in-transit points being classified as noise points and rejected from the final light curve, as illustrated in Figure\,\ref{fig:over_rej}.

While this could serve as a fast way to reliably remove large signals from a light curve and re-search the residuals for additional periodic signals, it could also have adverse
effects on a survey looking for transiting candidates.

Dealing with the issue of large signals requires an assumption and a caveat. The assumption (in our opion, quite justified) is that \emph{any} sufficiently large signal (of the order
of 3\% and more) will be readily detectable with transit-detection algorithms, such as the \texttt{BLS} algorithm \citep{kovacs1}, on the \emph{detrended} light curve itself, without
the need for any additional clipping or filtering. The caveat is that the presence of a large signal is, of course, not known beforehand, so that a transit-detection algorithm needs
to run first.

In \texttt{TSARDI}, we implement a safeguard against over-rejection by setting a strict lower limit for the percentage of points in the final light curve compared to the original
number of points, that is, $NPTSLIM=N_{\rm{fin}}/N_{\rm{org}}$. Depending on the tentative depth returned by the transit-detection algorithm, $NPTSLIM$ is set to 82\% for signals
up to 2\%; 95\% for signals up to 5\%; and 98\% for signals larger than 5\%.

\begin{figure}
 \includegraphics[width=9cm]{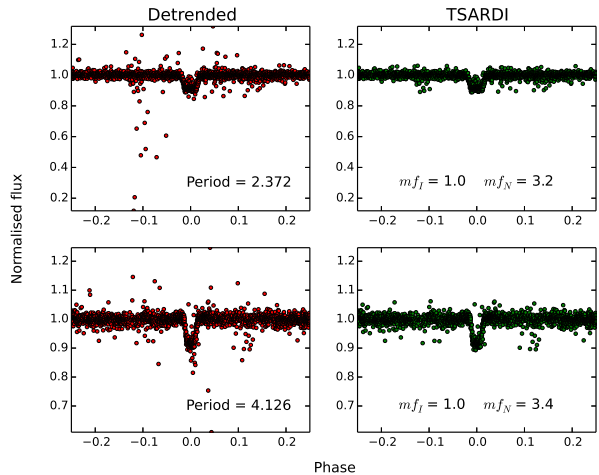}
 \caption{Two examples of transits of a 2\,$\rm{R_{J}}$ object (with orbital periods indicated), yielding depths of 0.088 and 0.132 (top and bottom, respectively). We show the
 detrended light curve on the left-hand side panels; and the final \texttt{TSARDI} filtered light curves on the right-hand side panels. We also indicate the corresponding
 $\left(mf_I,mf_N\right)$ values.}
 \label{fig:mfimfn}
\end{figure}

The $NPTSLIM$ is also the reason for the presence of the multiplication factors $mf_I$ and $mf_N$ first mentioned in Sec.\,\ref{subsubsec:cap}, as they allow for easy 
adjustment of the $Eps$ value at each Step of the core algorithm, depending on the current rejection rate compared to $NPTSLIM$. Varying the $\left(mf_I,mf_N\right)$ values
allows \texttt{TSARDI} to retain deep signals, as indicated in Figure\,\ref{fig:mfimfn}.

The interaction between all components is governed by the external-shell part of \texttt{TSARDI}, detailed below.

\subsubsection{The external-shell part}
\label{subsubsec:esp}
The external-shell part serves as a wrapper for the core-algorithm part. At its centre, it hosts a double loop designed to run consecutive iterations of the core-algorithm 
part, while safeguarding against over-rejection. The external-shell (and by extension, \texttt{TSARDI} itself) runs in the following fashion:

\begin{itemize}
 \item [$\bullet$] Two sets of values are defined for the multiplication factors $mf_I$ and $mf_N$; both sets range from 1 to 5, but with steps of 1 and 0.1 for $mf_I$ and $mf_N$
 respectively.
 \item [$\bullet$] \texttt{BLS} is run on the detrended light curve, and $NPTSLIM$ is set according to the result.
 \item [$\bullet$] A \emph{first} loop begins for each $mf_I$ value.
 \item [$\bullet$] A \emph{second} loop begins for each $mf_N$ value.
 \item [$\bullet$] For the given pair of $\left(mf_I,mf_N\right)$ values, the core algorithm runs on the detrended light curve.
 \item [$\bullet$] The percentage of remaining points ($PRP$) in the output light curve is recorded and compared against $NPTSLIM$.
 \item [$\bullet$] If $PRP\geq NPTSLIM$ both loops break, otherwise the algorithm continues with the next pair of $\left(mf_I,mf_N\right)$ values.
\end{itemize}

If the loops reach their end (i.e. $PRP<NPTSLIM$ for every pair of $\left(mf_I,mf_N\right)$ values), then the maximum recorded value of $PRP$ is compared to $NPTSLIM$. If the 
difference is less than 0.5\%, \texttt{TSARDI} is re-run with that pair of $\left(mf_I,mf_N\right)$ values giving $max\left(PRP\right)$; else, the detrended light curve remains
untouched.

Finally, we should note that we have settled on these specific choices for the values of different variables, such as the range and step of $mf_I$ and $mf_N$, the values for 
NPTSLIM and the Step-unique values of MinPTs, for consistently yielding the best results based on extensive tests carried out on data by the Qatar Exoplanet Survey (QES, 
\citealt{alsubai13}). Some adjustment might be required to these parameters and/or the Steps themselves when applying \texttt{TSARDI} to different sets of data; for example, splitting
the light curve into night segments isn't really applicable on continuous space-based data sets (but, perhaps, splitting into \emph{some} form of segments is).

\section{Results}
\label{sec:results}
To assess the performance of the algorithm, we selected a field from the QES, observed with one of the 400\,mm lenses (f/2.8,
FOV $5.24^{o}\times5.24^{o}$). This particular data set was collected over a period of two years, from Jan. 2013 to Jan. 2015, and consists of $\sim$4\,500 points, with an 
exposure time of 60\,sec. The data were reduced with the QES pipeline, described in detail in \cite{alsubai13}.

We limited the sample by imposing a cut on stellar magnitude of V < 14, resulting in 2022 stars. Following a similar procedure to the one described in \cite{collier}, for each star, 
we used the available $V$ and $K$ magnitudes, together with theoretical (and/or empirical) colour-temperature, temperature-radius and mass-radius relations to obtain a first estimate 
of the stellar masses and radii.

Subsequently, we injected a simulated transit signal of an $R_{P}=1.0\,\rm{R_{J}}$ planet, generated using the \cite{pal} model, in all the \emph{raw} light curves of the sample.
We did this for two different orbital periods $P_{1}\,=\,2.37217\,$d and $P_{2}\,=\,4.12669\,$d. The transit ephemeris was chosen so that an adequate number of transits would 
be sampled in the given data set, to ensure a large number of detections for statistical purposes. 

Subsequently, we detrended the light curves using the \texttt{DOHA} algorithm \citep{mislis17} and processed them further with \texttt{TSARDI}; we will refer to these light 
curves as the \emph{detrend \& \texttt{TSARDI}}, DT group. We also created a ``control'' group, the DC group, by processing the detrended light curves with a more general sigma-clip
algorithm, rejecting (i) points that were more than $8\sigma$ from the overall light curve mean; (ii) points that were more than $8\sigma$ from individual nightly means; and (iii)
nights whose standard deviation was more than $5\sigma$ from the average standard deviation.

\subsection{Overall RMS improvement}
\begin{figure}
 \includegraphics[width=9cm]{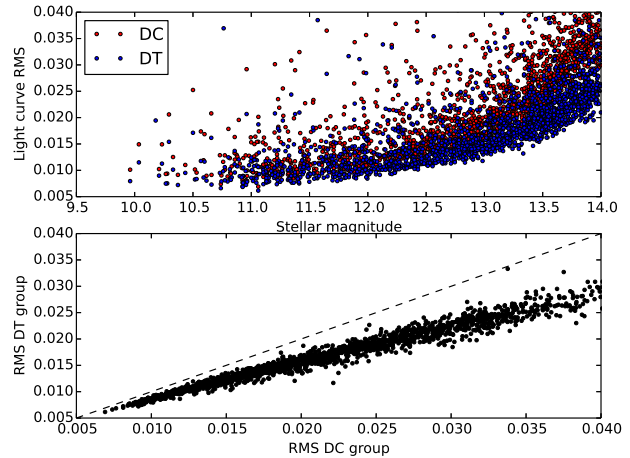}
 \caption{\textbf{Top panel:} RMS diagram for the DC (red) and the DT (blue) groups; see text for definition of the groups. \textbf{Bottom panel:} DC RMS versus DT RMS. In both
 panels, the overall RMS improvement is obvious, particularly for magnitudes below ~12.0.}
 \label{fig:rms}
\end{figure}

The first test illustrating the efficiency of our algorithm is a straightforward comparison of the light curve RMS between the DC and DT groups. In the top panel of 
Figure\,\ref{fig:rms} we plot a typical RMS diagram for both groups. The overall RMS improvement is obvious, indicating that our algorithm not only clips obvious outliers (the
scattered points in the upper left diagonal), but also that the additional steps of filtering out nights with comparatively high RMS can indeed improve the overall RMS of the main 
locus. This RMS improvement becomes more evident in the lower panel of Fig.\,\ref{fig:rms}, where we plot the RMS of the DC group versus that of the DT group. %Finally, we note that
%on average (over the entire sample) the algorithm classifies a $13\%\pm4\%$ of the total light curve points as noise points.

\subsection{Transit detection efficiency}
\begin{figure}
 \includegraphics[width=9cm]{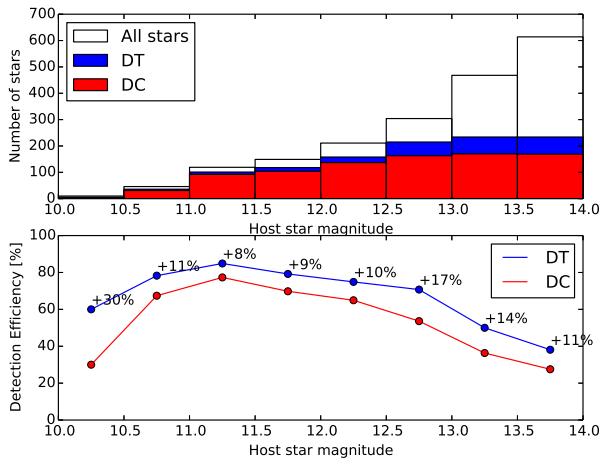}
 \caption{\textbf{Top panel:} Histogram of successful detections for the DC (red) and the DT (blue) groups per magnitude bin of the entire sample (white); 
 \textbf{Bottom panel:} Transit detection efficiency per bin. The numbers above the DT points indicate the difference in efficiency from the corresponding DC points.}
 \label{fig:histo}
\end{figure}

The major test for our algorithm was to investigate whether it can indeed (positively) affect the transit detection efficiency. For that, each light curve was subjected to the
\texttt{BLS} algorithm \citep{kovacs1}; this was done for both the DC and the DT groups. As ``successful recovery'', we consider the identification of the input planet period as
the dominant peak in the BLS periodogram. The results for $P_{1}\,=\,2.37217\,$d are shown in Figure\,\ref{fig:histo}.

In the top panel of Fig.\,\ref{fig:histo} we plot a histogram of the entire stellar sample in bins of 0.5 mag, together with the successful BLS detections in both the DC and the DT 
groups. It is clear that the DT detections are more than the DC ones, in every bin. To quantify the improvement, we plot the detection efficiency in each magnitude bin in 
the lower panel of Fig.\,\ref{fig:histo}, where the numbers correspond to the actual percentage difference between the two groups in each bin. 

In terms of absolute numbers, out of the possible 2022 planets, the DC group had a 45.3\% overall success rate (916 planets) versus a 56.5\% rate for the DT group (1142 
planets). There were 22 unique detections for the DC group, and 248 unique detections for the DT group, resulting in a very favourable 11:1 ratio for the latter. 
Selecting the subsample of the moderately bright end (V < 12.0), out of the maximum 329 planets, the DC group had a success rate of 73.3\% (241 planets), while the DT group had a 
success rate of 81.5\% (268 planets). For the fainter end (V > 12.0), where the RMS improvement with \texttt{TSARDI} becomes readily obvious (see again Fig.\,\ref{fig:rms}), out of 
the maximum 1693 planets, the DC group had a success rate of 39.9\% (675 planets), while the DT group successfully identified more than half the planets, with a 51.6\% success rate 
(874 planets).

\begin{figure}
 \includegraphics[width=9cm]{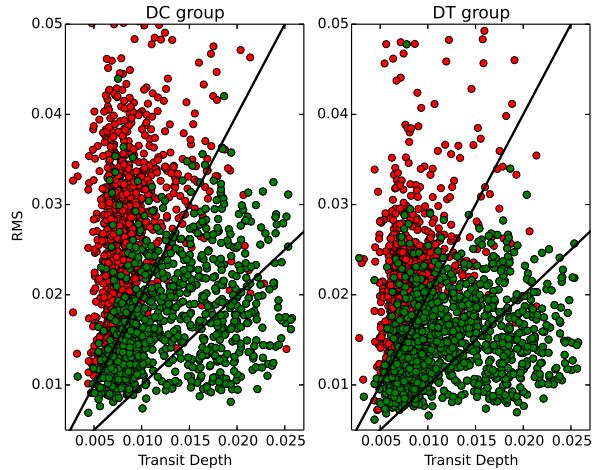}
 \caption{Transit depth versus RMS for the DC (left panel) and the DT (right panel) groups; successful and unsuccessful detections in both groups are plotted as green and red points,
 respectively. To aid the eye, we also plot the $D=RMS$ and $D=2*RMS$ lines. The vast majority of planets not detected have $D<1\%$ and $RMS>2*D$, and correspond to the fainter end
 ($V>13.0$) of the stars in our sample.}
 \label{fig:depth}
\end{figure}

For a more detailed look into the workings of the algorithm, we plot in Figure\,\ref{fig:depth} the expected transit depth $D$ (based on our initial estimate of $R_{*}$) versus RMS 
for both the DC and the DT groups, differentiating between successful and unsuccessful detections. We also plot the $D=RMS$ and $D=2*RMS$ lines. It is evident that the majority of
unsuccessful detections have small depth ($D<1\%$) and large RMS ($RMS>2*D$); most of these stars have $V>13.0$ and the photometric accuracy of the survey itself becomes the dominant
factor. Notice again the improvement in light curve RMS and how much ``tighter'' the RMS of the DT group becomes.

\begin{figure}
 \includegraphics[width=9cm]{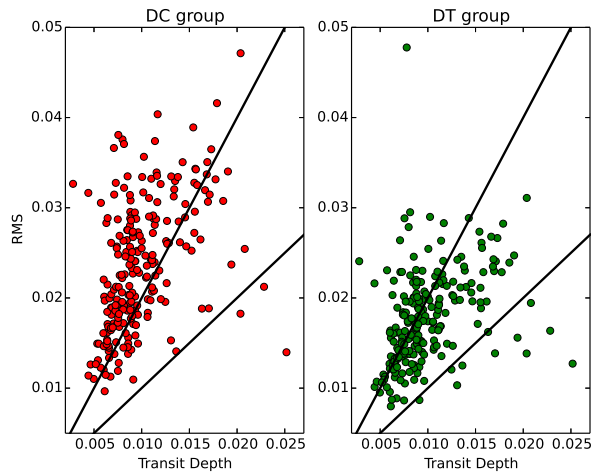}
 \caption{Same as Fig.\,\ref{fig:depth}, but collecting only the unique \texttt{TSARDI} detections. Note again the RMS improvement, and that the majority of these systems have 
 transit depths smaller than 1\%.}
 \label{fig:unidet}
\end{figure}

To further illustrate the difference between the more ``generic'' sigma-clip algorithm and \texttt{TSARDI}, we one again plot in Figure\,\ref{fig:unidet} the expected transit 
depth $D$ versus the light curve RMS, but this time, only for the 248 unique \texttt{TSARDI} detections. The ability of \texttt{TSARDI} to improve the overall RMS and pick up small
signals (the majority of unique detections have transit depths less than 1\%) is evident.

The results for the ssecond orbital period, $P_{2}\,=\,4.12669\,$d, are very similar, both qualitatively and quantitatively. The DC group had an overall success rate of
40.9\% versus 51.9\% for the DT group; for the moderately bright subsample, the success rates were 71.7\% versus 77.5\% for the DC and DT groups respectively; and finally the 
unique detection ratio was 9:1 in favour of the DT group (252 versus 29 planets).

\section{Conclusions}
\label{sec:conc}
We have developed \texttt{TSARDI}, a time-series analysis algorithm aiming to identify and remove residual data irregularities that remain in light curves, even after a detrending 
process. \texttt{TSARDI} is built on the clustering algorithm \texttt{DBSCAN}, and uses the latter's density-based notion, via an appropriately selected set of distance functions, to 
find outlying noise points; in our implementation, these noise points can be both ``traditional'' individual-point outliers, as well as individual \emph{nights} that are distinct and 
differ significantly from the majority of nights in a long-term light curve.

Based on the results of a search for (simulated) transits on a real data set, we demonstrate that \texttt{TSARDI} can lead to a substantial improvement of the transit detection 
efficiency; compared to light curves filtered with a straightforward sigma-clip algorithm, \texttt{TSARDI}-processed light curves showed an overall increase of $\sim$10\% in the 
number of detections. Taking into account the accuracy of the data used, and limiting the sample in terms of host star magnitude ($m<12$), leads to a detection rate of 80\% after 
using \texttt{TSARDI}.

\texttt{TSARDI} was conceived and tailor-built to deal with light curves from ground-based, large-scale surveys of transiting exoplanets. However, due to the flexibility of 
\texttt{DBSCAN}'s density-based clustering, and with appropriate choices for the key algorithm parameters, it can be easily adapted and extended to essentially any time series.

\section*{Acknowledgments}
We thank the anonymous referee for insightful comments and suggestions, which improved not only the original manuscript, but the algorithm itself as well. This publication was made 
possible by NPRP grant $\sharp$ X-019-1-006 from the Qatar National Research Fund (a member of Qatar Foundation). The statements made herein are solely the responsibility of the 
author.

\bsp

\label{lastpage}
\end{document}